\documentclass[12pt]{article}
\usepackage{feynmp,slashed}
\usepackage{amsmath}
\usepackage{amssymb}
\usepackage{amsthm}
\usepackage{psfrag}
\usepackage{graphicx}
\usepackage{hyperref}

\newcommand\bw{\begin{widetext}}
\newcommand\ew{\end{widetext}}
\newcommand{\be}{\begin{equation}}
\newcommand{\ee}{\end{equation}}
\newcommand{\beqa}{\begin{eqnarray}}
\newcommand{\eeqa}{\end{eqnarray}}

\newcommand{\pd}{\partial}
\newcommand\m{\mu}
\newcommand\D{\Delta}
\newcommand\n{\nu}

\renewcommand\a{\alpha}
\renewcommand\b{\beta}
\renewcommand\l{\lambda}

\newcommand\vk{\varkappa}

\def\e{{\rm e}}
\def\d{\partial}
\newcommand{\bseq}{\begin{subequations}}
\newcommand{\eseq}{\end{subequations}}

\begin{document}

\title{\vspace{-2cm} 
\begin{flushright}
{\normalsize
CERN-PH-TH-2014-191} \\
\vspace{-0.5cm}
{\normalsize INR-TH/2014-021}
\end{flushright}
\vspace{0.5cm} 
\bf Completing 
Lorentz violating massive gravity\\ at
high energies\footnote{Prepared for a special issue of JETP dedicated
to the 60th birthday of Valery Rubakov.}}

\author{Diego Blas$^1$,  Sergey Sibiryakov$^{1,2,3}$\\[2mm]
{\normalsize\it  $^1$ CERN Theory division, 
CH-1211 Geneva 23, Switzerland}\\[1.5mm]
{\normalsize\it $^2$Institut de Th\'eorie des Ph\'enom\`enes Physiques, 
 EPFL, CH-1015 Lausanne, Switzerland}\\[1.5mm]
{\normalsize\it $^3$ Institute for Nuclear Research of the
Russian Academy of Sciences,}\\[-0.05cm]
{\normalsize\it 60th October Anniversary Prospect, 7a, 117312
Moscow, Russia}}
\date{}
\maketitle

\begin{flushright}
\begin{minipage}{4.5cm}
{\em To Valery Rubakov, \\ Teacher and Colleague}
\end{minipage}
\end{flushright}
%\vspace{0cm}
\begin{abstract}
%\noindent
Theories with massive gravitons are interesting for a variety of
physical  applications, ranging from cosmological 
phenomena to holographic modeling of condensed matter systems. To date,
they have been formulated as effective field theories with a
cutoff proportional to a positive power of the graviton mass $m_g$ and  
much smaller than that of the massless theory ($M_P\approx 10^{19}\,
\mathrm{GeV}$ in the 
case of general relativity).
In this paper we present an ultraviolet completion for massive gravity
valid up to a high energy scale independent of 
the graviton mass. 
The construction is based on the existence of a
preferred time foliation combined with spontaneous condensation of vector fields.
The perturbations of
these fields are massive and below their mass
the theory 
reduces to a model of Lorentz violating massive gravity. 
The latter theory possesses instantaneous modes whose consistent
quantization we discuss in detail.
We briefly study some modifications to gravitational phenomenology at
low-energies. The homogeneous cosmological solutions are the same as
in the 
 standard cosmology. 
The gravitational potential of point sources agrees with the Newtonian one at 
distances small with respect to $m_g^{-1}$. 
Interestingly, it becomes repulsive at 
larger distances.
\end{abstract}

\newpage

\tableofcontents
%%%%%%%%%%%%%%%%%%%%%%%%%%%%%%%%
\section{Introduction}
%%%%%%%%%%%%%%%%%%%%%%%%%%%%%%%%

Can gravity  be mediated by a  massive tensor field ? This
straightforward question has generated  a lot of controversy 
since it was first formulated by Fierz and Pauli
\cite{Fierz:1939ix}. 
The situation is remarkably different from the case of gauge
interactions mediated by vector fields, where the Higgs mechanism
provides a clear-cut way to give mass to the vector bosons within a
weakly coupled theory. The differences fall into two categories. 
First, a generic Lorentz invariant theory with massive spin-2 fields
(gravitons) presents instabilities in the sector of additional
polarizations appearing in the massive, as opposed to massless, case
--- the ``Goldstone'' sector. These instabilities arise 
around realistic backgrounds and endanger the consistency of the theory
even at low energies
\cite{Boulware:1973my,Creminelli:2005qk}. It was first realized by
V.~Rubakov \cite{Rubakov:2004eb} that these problems can be avoided by
breaking the Lorentz invariance. This approach has lead to the
formulation of a class of Lorentz violating (LV) massive gravities as
consistent effective field theories (EFTs) \cite{Dubovsky:2004sg} (see
\cite{Rubakov:2008nh} for review). An alternative way to improve the
behavior of the Goldstone sector while preserving the Lorentz
invariance has been found in \cite{deRham:2010kj} and consists in a
judicious choice of the couplings for the interactions of the massive
gravitons (see e.g. \cite{Hinterbichler:2011tt,deRham:2014zqa}) for
reviews). 
It has been argued \cite{deRham:2012ew,deRham:2013qqa} 
that this tuning might be stable under quantum
corrections, but it is not clear at the moment if there is any
symmetry behind.

The second difference between massive spin-2 and spin-1 theories lies
in their different behavior at high energies. The interactions in the
Goldstone sector of massive gravity become strong and the perturbation
theory
breaks down at a certain  cutoff scale\footnote{Throughout the paper
  we identify the cutoff with the strong coupling scale of the
  perturbation theory around the Minkowski background. In the setup
  of \cite{deRham:2010kj} the scale of strong coupling may be raised
  in curved space-time due to the Vainshtein mechanism (see the
  discussion in \cite{deRham:2014zqa}). However, the validity 
of the corresponding backgrounds is under debate
  \cite{Kaloper:2014vqa}. }
$\Lambda_{\rm low}$ depending on the graviton mass $m_g$. In the limit
of vanishing mass this scale goes down to zero and, instead of
recovering the massless case, the theory ceases to exist. In
fact, the same is also true for a pure theory of massive vector
fields with non-abelian interaction. However, in the latter case the
ultraviolet (UV) completion is known in the 
form of the Higgs mechanism which makes the theory renormalizable by
adding a handful of new degrees of freedom (Higgs
bosons). Importantly, in the resulting theory the massless limit is
perfectly smooth and corresponds to the restoration of the
spontaneously broken gauge symmetry\footnote{From a purist's
  viewpoint, no symmetry is broken in the Higgs mechanism, gauge
  invariance being just a redundancy in the description. However, we
  allow ourserves
  this abuse of terminology because of its clear intuitive
  meaning.}. 

No such mechanism has been found so far for massive
gravity\footnote{Exceptions are theories in AdS where the mass of the
  graviton can appear due to non-trivial conditions at the
  time-like
  boundary~\cite{Porrati:2001db,Kiritsis:2006hy,Aharony:2006hz}. 
However, these constructions rely heavily on the
peculiar properties of the AdS geometry. In particular, the resulting
graviton mass is always parametrically smaller than the inverse
AdS radius.}.     
Of course, in this case one cannot insist on an embedding into a fully
UV complete theory --- the massless theory being
non-renormalizable anyway with the cutoff at the Planck mass
$M_P\approx 10^{19}$GeV (see, however, Sec.~\ref{sec:beyondphi0}). 
Still, it makes sense to search for a setup,
whose cutoff would be independent of the graviton mass and as close to
the Planck scale as possible. To preserve the analogy with the Higgs
mechanism, the embedding theory must contain only a finite number of
new degrees of freedom compared to massive gravity 
itself\footnote{This excludes the known theories with massive spin-2
  fields, such as Kaluza--Klein models or string theory: both 
 imply the presence of an infinite tower of new degrees
  of freedom with masses of order $\Lambda_{\rm low}$.}, these degrees of
freedom must be weakly 
coupled and the limit $m_g\to 0$ must be regular. The goal of this
work is to present a setup fulfilling the above requirements. 

The reasons for pursuing this endeavor are not merely
academic. First, massive gravity is a very natural candidate for an
infrared (IR) modification of general relativity (GR) \cite{Blas:2008uz}. 
Such modifications are an interesting  playground to look for
alternatives to the cosmological constant as the source of 
cosmic acceleration (see e.g. \cite{Joyce:2014kja}).  
The acceleration may be generated
at the level of the background 
(new contribution to the Friedmann equation) 
or of perturbations (weaker or repulsive gravitational
potential at distances larger than $m_g^{-1}$).  
It is fair to say that  none of these possibilities have been
satisfactory implemented so far in concrete models.   
In any event, any candidate to explain cosmic acceleration should have a
completion at high energies, which is important for predictions
related to the early universe or very dense objects.  
This can also allow to understand the relevance of certain
tunings of the IR  parameters. 
We will see how a concrete ultraviolet (UV) completion 
can have nontrivial consequences at very large distances. 

Massive gravity has also been discussed in the context of the
gauge/gravity correspondence \cite{Kiritsis:2006hy,Aharony:2006hz}.  
Various phases of massive gravity may be useful
to describe different phenomenology at strong coupling.  
In particular, it was recently realized that LV
massive gravity is relevant
for the description of systems with broken
translational invariance
\cite{Vegh:2013sk,Blake:2013owa}. 
The completion of the theory to smaller distances on the gravity side
yields access to the operators of higher dimensions in the strongly
coupled field theory\footnote{We thank Riccardo Rattazzi for the
  discussion of this point.}.

Finally, the theory of massive gravity is related to the spontaneous
breaking of space-time symmetries
\cite{ArkaniHamed:2002sp,Dubovsky:2004sg}. This is an appropriate 
language to describe different states of matter within the EFT framework
\cite{Son:2005ak,Dubovsky:2005xd}. One can speculate that 
the Higgs mechanism for massive gravity will be relevant to describe the
phase transitions in such systems.

In this paper we will focus on the LV massive gravity of 
\cite{Rubakov:2004eb,Dubovsky:2004sg}. 
Our main motivation for this choice is the already mentioned validity
of this theory as a low-energy EFT, whose structure is protected by
symmetries. Besides, the fundamental role of Lorentz invariance 
in quantum gravity has
been questioned recently \cite{Horava:2009uw}.
It is interesting to explore if massive gravity can be naturally
embedded in this framework\footnote{See \cite{CuadrosMelgar:2011yw}
  for an early 
  attempt in this direction.}.   
We will assume that at the fundamental level the violation of 
Lorentz invariance is minimal and amounts to 
the existence
of a preferred foliation of space time
\cite{Horava:2009uw,Blas:2010hb}.  
It is worth noting that the presence of superluminal propagation 
\cite{Gruzinov:2011sq,Burrage:2011cr,Deser:2014hga} 
in the seemingly Lorentz invariant massive gravity of
\cite{deRham:2010kj} makes a Lorentz invariant
Wilsonian UV completion of this theory problematic 
\cite{Adams:2006sv}. Thus, even in this case the UV
completion (if any) is likely to be Lorentz violating.

The paper is organized as follows. In Sec.~\ref{sec:LV} we briefly 
summarize the formalism of LV massive gravity and define
the phase that we will consider. In Sec.~\ref{sec:beyond} we  
introduce the UV completion that allows to push the cutoff of the
theory to values close to $M_P$. We analyze the background
solutions of the theory in Sec.~\ref{sec:cosmology} and 
show that the LV massive gravity of Sec.~\ref{sec:LV} appears in
the IR limit. 
Section~\ref{sec:hier} is devoted to the analysis of the  
degrees of freedom in the theory at different scales. We also discuss
in some detail the quantization of instantaneous modes present in the
LV massive gravity and their relation to a certain type of
non-locality along the spatial directions. 
First results in phenomenology are presented in
Sec.~\ref{sec:Newton}. We conclude with the summary and discussion
in Sec.~\ref{sec:discussion}.

%%%%%%%%%%%%%%%%%%%%%%%%%%%%%%%%
\section{Lorentz violating massive gravity}
%%%%%%%%%%%%%%%%%%%%%%%%%%%%%%%%
\label{sec:LV}

We will now briefly review the construction of LV massive gravity. We
will formulate these theories in a language closer to the symmetry
breaking mechanism by introducing  St\"uckelberg fields
\cite{ArkaniHamed:2002sp,Dubovsky:2004sg}. This
formulation is useful to understand many features of the theories, in
particular the strong coupling scale.

We focus on the setup where Lorentz invariance is broken down to the
subgroup of spatial rotations \cite{Rubakov:2004eb}.   
To describe this situation, let us consider four scalar
St\"uckelberg fields, 
$\phi^0$, $\phi^a$, $a=1,2,3$, with internal $SO(3)$ symmetry acting on the
indices $a$, coupled to the metric in a covariant way. Additional
symmetries must be imposed on this sector to protect it from
pathologies \cite{Dubovsky:2004sg}. We start by requiring invariance
under the shifts\footnote{We start from the simple $\phi^0$-shifts to
  make contact with \cite{Dubovsky:2004sg}. Later on we will promote
  them to a larger symmetry, see Eq.~(\ref{repar}).} of $\phi^0$,
\bseq
\label{phishifts}
\be
\label{phi0shifts}
\phi^0\mapsto \phi^0+const\;,
\ee
and the $\phi^0$-dependent
shifts of $\phi^a$,
\be
\label{phiashifts}
\phi^a\mapsto \phi^a+f^a(\phi_0)\;,
\ee  
\eseq
where $f^a$ are arbitrary functions. We assume that in the stationary
state  
the St\"uckelberg fields acquire
coordinate-dependent vacuum expectation values (VEVs), 
\be
\label{VEVs}
\phi^0=\mu_0 ^2\,t~,~~~\phi^a=\mu^2 x^a\;.
\ee
These VEVs break the product of 4-dimensional diffeomorphisms and the
internal symmetries of St\"uckelberg fields down to the diagonal
subgroup consisting of the time shifts,
\bseq
\label{resdiffeos}
\be
\label{resdiffeos1}
t\mapsto t+const\;,
\ee
time-dependent shifts of the spatial coordinates,
\be
\label{resdiffeos2}
x^a\mapsto x^a+f^a(t)
\ee
\eseq
and $SO(3)$ spatial rotations. A simple Lagrangian that obeys the
imposed symmetries and admits the VEVs (\ref{VEVs}) 
has the form\footnote{This Lagrangian is not the most general one, but it is
  sufficient for our purposes, as it reproduces all possible mass
  terms for the metric which are allowed by the symmetries (\ref{phishifts}).}
${\cal L}_S={\cal L}_{S1}+{\cal L}_{S2}$,
\bseq
\label{Lsimple}
\begin{align}
\label{Lsimple1}
&{\cal L}_{S1}=\frac{1}{8\mu_0^4}\big((\d_\m\phi^0)^2-\mu_0^4\big)^2
-\frac{\kappa_0}{4\m_0^4}\big((\d_\m\phi^0)^2-\mu_0^4\big)
\big(\sum_aP^{\m\n}\d_\m\phi^a\d_\n\phi^a+3\m^4\big)\;,\\
\label{Lsimple2}
&{\cal L}_{S2}=
-\frac{1}{8\m^4}\sum_{a,b}
\big(P^{\m\n}\d_\m\phi^a\d_\n\phi^b+\m^4\delta^{ab}\big)^2
+\frac{\kappa}{8\m^4}
\big(\sum_a P^{\m\n}\d_\m\phi^a\d_\n\phi^a+3\m^4\big)^2\;, 
\end{align}
\eseq
where $\kappa_0$, $\kappa$ are dimensionless constants and
\be
\label{covdiv}
P_{\m\n}=g_{\m\n}-\frac{\d_\m\phi^0\d_\n\phi^0}{g^{\l\rho}\d_\l\phi^0\d_\rho\phi^0},
\ee 
is the projector on the subspace orthogonal to the gradient of
$\phi_0$ which ensures invariance under (\ref{phiashifts}).

To understand the physical content of the theory, let us consider
small fluctuations of the metric and expand to the quadratic order in 
$h_{\m\nu}\equiv g_{\m\n}-\eta_{\m\n}$.
Using general covariance, we can identify the coordinates with the
St\"uckelberg fields, as in (\ref{VEVs}). In other words, we work in
the gauge, where the fields $\phi^0$, $\phi^a$ do not fluctuate; we
call it ``unitary gauge''. Then,
the quadratic Lagrangian takes the form,  
\be
\label{h1}
{\cal L}_S^{(2)}=\frac{\m_0^4}{8} h_{00}^2-\frac{\m^4\kappa_0}{4} h_{00} h_{aa}-
\frac{\m^4}{8}h_{ab}h_{ab}+\frac{\m^4\kappa}{8} h_{aa}^2\, ,
\ee
where the summation over repeated indices is understood. This is
precisely a mass term for the metric perturbation. In particular, 
the graviton (the helicity-2 component) acquires the mass 
\be
\label{gravmass}
m_g=\frac{\m^2}{M_P}\;,
\ee
where $M_P$ is the Planck mass.
Note that the term $h_{0a}h_{0a}$ which is missing in (\ref{h1})
compared to the most general expression \cite{Rubakov:2004eb} is
forbidden by the residual symmetry (\ref{resdiffeos2}).  
The quadratic Lagrangian of the form (\ref{h1})
appears also in bimetric theories 
\cite{Berezhiani:2007zf,Blas:2007ep}.
 
Let us return from the unitary gauge to the covariant Lagrangian
(\ref{Lsimple}) which is more suitable to study the non-linear
properties. Importantly, in the case when $\m_0$, $\m$ are much
smaller than $M_P$ we can decouple the metric fluctuations and
concentrate on the St\"uckelberg fields, as if they were living in
flat space-time. We write
\be
\label{Stexp}
\phi^0=\m_0^2t+\psi^0~,~~~~\phi^a=\m^2x^a+\psi^a\;
\ee
and obtain
\be
\label{psi1}
{\cal  L}_S=\frac{(\dot\psi^0)^2}{2}
+\frac{\kappa_0\m^2}{\m_0^2}\dot\psi^0\d_a\psi^a
-\frac{\d_a\psi^b\d_a\psi^b}{4}
-\frac{(1-\kappa)}{4}(\d_a\psi^a)^2
+{\cal L}_{int}\bigg(\frac{\d\psi^0}{\m_0^2},\frac{\d\psi^a}{\m^2}\bigg)
\;,
\ee
where the last term stands for the derivative interactions of cubic
and higher orders. By power-counting, the strength of these
interactions grows with the increase of energy or momentum and the
theory breaks down at the scale   
\be
\label{cutoff}
\Lambda=\min\{\m_0,\m\}\;.
\ee
Comparing with (\ref{gravmass}) we conclude that the cutoff is bounded
from above,
\be
\label{cutoffup}
\Lambda < \Lambda_2\equiv\sqrt{m_g M_P}\;.
\ee
Actually, this conclusion is not related to the specific form of the
Lagrangian (\ref{Lsimple}). As discussed in  \cite{Dubovsky:2004sg},
$\Lambda_2$ provides an absolute upper bound on the cutoff in a
general massive gravity theory formulated in terms of the metric and
St\"uckelberg fields only\footnote{The cutoff is even lower,
  $\Lambda<\Lambda_3\equiv(m_g^2 M_P)^{1/3}$, if one restricts to the
  Lorentz invariant theories \cite{ArkaniHamed:2002sp,Schwartz:2003vj}.}. 
Thus the theory does not admit a smooth
limit $m_g\to 0$.

From the quadratic part of (\ref{psi1}) we read off that a linear
combination of $\psi^0$ and the longitudinal part of $\psi^a$ has
degenerate dispersion relation
\be
\label{degen1}
\omega^2=0\;.
\ee
This presents a potential danger, as in non-trivial backgrounds the
r.h.s. of the dispersion relation can become negative leading to an
instability. 
In \cite{Dubovsky:2004sg} it was suggested to lift the degeneracy by
adding to the Lagrangian quadratic terms with higher derivatives as in
the ghost condensate model \cite{ArkaniHamed:2003uy}.
In the next section we will take a different route and embed the field
$\phi^0$ into the 
 khronometric model \cite{Blas:2010hb}.
 
The rest of the modes in (\ref{psi1}) --- the transverse part of
$\psi^a$ and the longitudinal component of $\psi^a$ linearly
independent from $\psi^0$ --- obey the equations of the form,
\be
\label{degen2}
\bar{k}^2\psi=0\;,
\ee 
where $\bar{k}$ is the absolute value of the three-momentum $\bar k_i$.
Thus, for any non-zero spatial momentum these modes must vanish 
implying that there are no propagating degrees of freedom associated
with $\psi^a$. 
The symmetry (\ref{phiashifts}) ensures that this property is
preserved upon inclusion of higher-order operators 
\cite{Dubovsky:2004sg} and in curved backgrounds \cite{Blas:2009my}. 
Therefore the theories based on the symmetries
(\ref{phishifts}) present a class of well-defined EFTs\footnote{A subtle
issue of the proper treatment of the non-propagating modes at the
quantum level will be discussed in Sec.~\ref{sec:quantum}.}.

%%%%%%%%%%%%%%%%%%%%%%%%%%%%%%%%
\section{Going beyond $\Lambda_2$: ingredients}
%%%%%%%%%%%%%%%%%%%%%%%%%%%%%%%%
\label{sec:beyond}

%%%%%%%%%%%%%%%%%%%%%%%%%%%%%%%%
\subsection{The field $\phi^0$}\label{sec:beyondphi0}
%%%%%%%%%%%%%%%%%%%%%%%%%%%%%%%%

There are two natural ways to deal with the field $\phi^0$ to
complete the previous actions to energies higher than (\ref{cutoff}). 
First, as we
mentioned  
above, the degeneracy of the dispersion relation \eqref{degen1} can be
lifted by adding 
higher derivative terms as in the ghost condensate \cite{ArkaniHamed:2003uy}.
This theory is still an effective theory with a cutoff of order 
$\mu_0$, but since 
this is independent of the mass of the graviton, 
this scale can be quite high. Phenomenological bounds set the
constrain $\mu_0 \lesssim 10\,\mathrm{MeV}$. 
It was argued in \cite{ArkaniHamed:2005gu} that these bounds can be
relaxed by the non-linear dynamics which may push the upper limit on
$\m_0$ to $100\,\mathrm{GeV}$. This still remains much below the
Planck scale. The way to raise the cutoff of the theory to (almost)
Planckian was proposed in \cite{Ivanov:2014yla}. It uses the embedding
of the ghost condensate into the khronometric model \cite{Blas:2010hb}
and requires the introduction of a new degree of freedom --- khronon ---
at a scale below $\m_0$. One could use this strategy
here. 

However, it is more economical to use the second option and
identify the St\"uckelberg field 
$\phi^0$ directly with the khronon, thus keeping only a
single degree of freedom in this sector. In this case the symmetry
(\ref{phi0shifts}) is extended to the full reparameterization invariance, 
\be
\label{repar}
\phi^0\mapsto f^0(\phi^0)\;,
\ee
for an arbitrary monotonic function $f^0$. This larger symmetry forbids the
terms present in (\ref{Lsimple1}). Instead, the Lagrangian must be constructed
using the unit vector
\be
\label{um}
u_\m\equiv\frac{\d_\m\phi^0}{\sqrt{g^{\l\rho}\d_\l\phi^0\d_\rho\phi^0}}
\ee
invariant under the symmetry (\ref{repar}). The most general Lagrangian with the
lowest number of derivatives, and thus dominant at low energies, reads
\cite{Blas:2010hb},
\be
\label{Lkh}
{\cal L}_{kh}=-\frac{M_P^2}{2}\big(R+\b \nabla_\m u^\n\nabla_\n u^\m
+\l(\nabla_\m u^\m)^2+\a u^\m u^\n \nabla_\m u^\rho\nabla_\n u_\rho\big)\;,
\ee
where we have also included the standard GR action;  
$\a,\b,\l$ are dimensionless coupling constants. This can be 
combined with (\ref{Lsimple2}) to give an action of LV massive gravity.
Note that in the
unitary gauge\footnote{Because of the invariance (\ref{repar}) the
  choice of the constant $\m_0^2$ in the first formula of (\ref{VEVs})
is now arbitrary and unrelated to the parameters of the theory.} 
(\ref{VEVs}) the Lagrangian (\ref{Lkh})
 gives rise only to terms
with derivatives of the metric perturbations and thus does not
contribute to the mass term for $h_{\m\n}$. The latter 
reduces to 
\be
\label{h11}
{\cal L}_{\mathrm{mass}}=-\frac{\m^4}{8} h_{ab}h_{ab}+\frac{\m^4 \kappa}{8}
h_{aa}^2\;.
\ee
This can
be understood as the consequence of the time-reparameterization
invariance 
\be
\label{trep}
t\mapsto f^0(t)\;,
\ee  
which, being now a residual symmetry in the unitary gauge, forbids any
contributions containing $h_{00}$ without derivatives. This version of
massive gravity was considered in \cite{Gabadadze:2004iv}.

The Lagrangian (\ref{Lkh}) contains higher derivatives of the field
$\phi^0$ and one may be worried that this can lead to pathologies
(ghosts or gradient instabilities). In fact, this does not happen, as
the extra derivatives act in the spatial directions. This property
becomes explicit in the gauge where the time coordinate is identified
with $\phi^0$ as in the first equation of (\ref{VEVs}). Note, that
this identification still leaves the free choice of the spatial
coordinates, so it should not be confused with the unitary gauge where
{\em all} coordinates are fixed. We will call this partial gauge
fixing ADM gauge. The action of the khronometric model takes the form, 
\be
\label{khronoact}
S_{kh}=\frac{M_P^2}{2}\int dt\,
d^3x\,\sqrt{\gamma}N\bigg[(1-\b)K_{ij}K^{ij}-(1+\l)K^2
+{}^{(3)}R+\a\bigg(\frac{\d_iN}{N}\bigg)^2 \bigg]\;,
\ee 
where we used the Arnowitt--Deser--Misner (ADM) decomposition for the metric,
\be
\label{ADMmetr}
ds^2=dt^2-\gamma_{ij}(dx^i+N^idt)(dx^j+N^jdt)\;,
\ee
the extrinsic curvature of the constant-time slices
\begin{align}
K_{ij}=\frac{1}{2N}(\dot\gamma_{ij}-\nabla_iN_j-\nabla_jN_i)
~,~~~K=\gamma^{ij}K_{ij},
\end{align}
and denoted ${}^{(3)}R$ the three-dimensional curvature constructed from the
metric $\gamma_{ij}$. 
Apart from the symmetry (\ref{trep}), this action is invariant under time-dependent
spatial diffeomorphisms
\be
\label{spacediff}
x^i\mapsto\tilde x^i({\bf x},t)\;.
\ee
We will refer to the group 
consisting of the transformations (\ref{trep}) and (\ref{spacediff}) as
foliation-preserving diffeomorphisms (FDiff).
Clearly, the action (\ref{khronoact}) leads to equations of motion
which are second order in time derivatives. We will
work in the ADM gauge from now on.

The choice $\a=\b=\l=0$ corresponds to GR and the restoration of the
full dif\-feomor\-phisms-invariance. However, the limit $\a,\b,\l\to 0$ is
not smooth. At any non-zero values of 
$\a$, $\b$, $\l$ the theory propagates in addition to the helicity-2
gravitons a single helicity-0 mode (khronon). The latter has linear
dispersion relation; in the case $\a,\b,\l\ll 1$ (which is the relevant
one for phenomenology) it reads\footnote{This relation gets modified
  --- in particular, the khronon acquires a mass gap --- when the action
  (\ref{khronoact}) is coupled to the other sectors needed to
  reproduce the massive gravity, see Sec.~\ref{sec:Newton} below.} 
\cite{Blas:2010hb},
\be
\label{khronodisp1}
\omega^2=\frac{\b+\l}{\a}\;k^2\;.
\ee 
Due to the non-linear interactions of the khronon present in
(\ref{khronoact}) the model has a cutoff 
\be
\label{cutoffkh}
\Lambda_{kh}\sim M_P\min\{\sqrt{\a},\sqrt{\b},\sqrt{\l}\}\;.
\ee
Phenomenological considerations put upper bounds on $\a$,
$\b$, $\l$ \cite{Blas:2010hb,Shao:2013wga,Yagi:2013qpa} and hence constrain the
cutoff to be somewhat smaller than the Planck scale,
\be
\label{Lkhcutoff}
\Lambda_{kh}\lesssim 10^{15}\,{\rm GeV}\;.
\ee
Still, this is well above virtually any scale that can appear in
the astrophysical or cosmological context\footnote{In the applications
unrelated to astrophysics, such as non-relativistic holography or
description of solids, the parameters $\a$, $\b$, $\l$ are a priori
constrained only by the stability requirements, that are mild,
and the scale $\Lambda_{kh}$ can be as high as $M_P$.}. 
Furthermore, it is known how to complete the action (\ref{khronoact})
beyond $\Lambda_{kh}$ by embedding it into the Ho\v rava gravity  
\cite{Horava:2009uw,Blas:2009qj}. The latter presents a power-counting
renormalizable theory. However, due to the technical complexity,
 the question about its renormalizability
in the strict sense and UV behavior still remains open 
(see Refs.~\cite{Benedetti:2013pya,D'Odorico:2014iha} 
addressing this issue
in restricted settings). 
In these circumstances a cautious reader may prefer to take modest
attitude and view the khronometric model as an EFT with the cutoff
(\ref{cutoffkh}), which is sufficient for the purposes of this work.

Finally, let us mention the following peculiarity of the action
(\ref{khronoact}). As described in \cite{Blas:2010hb}, it leads to a
certain type of instantaneous interactions mediated by a
non-propagating mode. The latter is similar to the non-propagating
modes of massive gravity discussed in Sec.~\ref{sec:LV}. We will study
the instantaneous modes in more detail in Sec.~\ref{sec:quantum}. 

%%%%%%%%%%%%%%%%%%%%%%%%%%%%%%%%
\subsection{The fields $\phi^a$ and their coupling to Higgs vectors}
%%%%%%%%%%%%%%%%%%%%%%%%%%%%%%%%

Next we consider the triplet 
of St\"uckelberg fields invariant under
\be
\label{phiash1}
\phi^a\mapsto \phi^a+f^a(t)\;,
\ee
which is nothing but the symmetry (\ref{phiashifts}) in the ADM gauge.
We want a Lagrangian that admits the coordinate-dependent VEVs
(\ref{VEVs}), but at the same time is UV complete past the scale
$\m$. This precludes from introducing any self-interaction of the
St\"uckelberg fields involving the scale $\m$. Then the simplest option
is to choose the Lagrangian to be quadratic in $\phi^a$.  
To
respect the symmetry (\ref{phiash1}), it
 must depend only on the
{\em spatial} derivatives of these fields,
\be
\label{phiact}
S_\phi=\int dt\,d^3x\,\sqrt\gamma
N\bigg[-\frac{1}{2}\gamma^{ij}\d_i\phi^a\d_j\phi^a \bigg]\;.
\ee 
This does not
introduce any new strong coupling scale. 
However, without any further interactions this Lagrangian is not enough to
provide non-zero graviton mass. 
Though the configuration $\phi^a=\Phi x^a$ is a solution of the
  equations following from (\ref{phiact}) for any constant $\Phi$, it
  introduces non-vanishing energy density and pressure which make the
  universe expand\footnote{These density and pressure cannot be
    canceled by any bare cosmological constant.}. 
 As will become clear in the Sec.~\ref{sec:cosmology},
in this case the generated mass will decrease
with time and will  asymptotically vanish. 
Time varying masses can be interesting (see e.g. \cite{Blas:2009my}) but 
are not the aim of this paper. 
To give constant graviton mass, the VEVs in an expanding universe
must grow proportionally to the scale factor, $\phi^a\propto
a(t)x^a$, which is not a solution of the field equations implied by
(\ref{khronoact}) and (\ref{phiact}).
We have to add more ingredients.

Consider a triplet of vector fields with purely spatial components
$V^i_a$. These transform as vectors under the
diffeomorphisms preserving the foliation structure of the ADM gauge,
which act on the $i$-index. Besides, they form the fundamental
representation of a global internal
$SO(3)$  acting on  the index $a$. We do not assume any gauge
invariance associated to these vectors. To avoid strong coupling, we
focus on Lagrangians which are  
{\em renormalizable} 
in flat space-time. By the standard power-counting, they can contain
the derivatives of $V^i_a$ only quadratically and up to quartic terms
in the potential. The generic Lagrangian satisfying these properties
and invariant under  
$\mathit{FDiff}\times SO(3)$
reads
\be
\label{vectact}
\begin{split}
S_V=\int dtd^3x\,\sqrt\gamma
N\bigg[\frac{1}{2N^2}(\dot V^i_a&-N^j\nabla_j V^i_a+V^j_a\nabla_j N^i)^2
-\frac{c_1^2}{2}(\nabla_i V^j_a)^2-\frac{c_2^2}{2}(\nabla_i V^i_a)^2\\
&-\frac{\vk_1}{4}(V_a^iV_b^j\gamma_{ij}-M_V^2\delta_{ab})^2
-\frac{\vk_2}{4}(V_a^iV_a^j\gamma_{ij}-3M_V^2)^2\bigg]\;,
\end{split}
\ee
where $c_1$, $c_2$, $\vk_1$, $\vk_2$ are dimensionless couplings and
we have chosen the overall constant in the potential
to have vanishing vacuum energy (we will shortly introduce a
cosmological constant term in a different part of the action).
For clarity, 
we have omitted non-minimal interactions with the metric, such as 
${}^{(3)}R_{ij}V^i_a V^j_a$, ${}^{(3)}R V^i_a V^j_a\gamma_{ij}$, which
vanish in Minkowski space-time.
These terms would make the analysis more cumbersome without changing
it qualitatively.

When $M_V^2>0$ the vectors develop VEVs,
\be
\label{VVEVs}
V^i_a=M_V\delta^i_a\;,
\ee
which
break the product $SO(3)\times SO(3)$ of spatial and
internal rotations down to the diagonal subgroup,
cf.~\cite{Bento:1992wy,ArmendarizPicon:2004pm,Libanov:2005vu,Gorbunov:2005dd}. 
Below the scale $\sim
\sqrt{\vk} M_V$
the dynamics is described by the $\sigma$-model corresponding to this
pattern of symmetry breaking with the coset space defined
by
\be
\label{constr}
V_a^iV_b^j\gamma_{ij}=M_V^2\delta_{ab}\;.
\ee 
As the vector VEVs introduce an additional source of Lorentz symmetry
breaking, it is natural to expect that the phenomenological constraint on
the scale $M_V$ will be similar to that of 
$\Lambda_{kh}$, Eq.~\eqref{Lkhcutoff}. 
Notice, however, that $M_V$ is not related to the cutoff and can be
much lower than $\Lambda_{kh}$ without jeopardizing the validity of
the theory.

Finally, we complete our Lagrangian with a term mixing the vectors
and the St\"u\-ckel\-berg fields, 
\be
\label{SVphi}
S_{V\phi}=\int dt\,d^3x\,\sqrt\gamma
N \big[m_{A}V^i_a\d_i\phi^a-{\cal V}_0\big]\;.
\ee
This mixing operator has dimension 3 and thus is just a relevant deformation
of the previous action. It does not affect the UV properties of the
theory, in particular, it does not introduce any new UV cutoff, and
the parameter $m_{A}$ can be arbitrarily small without encountering any
singularity. We are going to see that in IR
this coupling leads to the
generation of the VEVs (\ref{VEVs}) with $\m^2=m_{A} M_V$ 
and the graviton mass (\ref{gravmass}).
The last term in (\ref{SVphi}) represents a cosmological constant and
can be tuned to cancel the negative vacuum energy that would be
generated otherwise (see below).
Note that it is technically natural to take 
the parameter $m_{A}$ to be much smaller than the other scales of the
theory as it is protected from large quantum corrections
by the
discrete symmetry\footnote{A similar argument is used to protect the
  small coupling between a time-like vector field and an ordinary
  massless scalar in the technically natural dark energy model of 
 \cite{Blas:2011en}.}  $\phi^a\mapsto -\phi^a$. 
In what follows we will assume the hierarchy of scales,
\be
\label{hierarchy}
M_P\gtrsim \Lambda_{kh}\gtrsim M_V\gg m_{A}\;. 
\ee
It is worth stressing that this hierarchy is not required by the
internal consistency of the theory. For example, one could consider
instead $m_{A}\sim M_V$. However, assuming (\ref{hierarchy}) makes
the physical picture particularly transparent.

%%%%%%%%%%%%%%%%%%%%%%%%%%%%%%%%
\section{Generation of VEVs in expanding backgrounds}
\label{sec:cosmology}
%%%%%%%%%%%%%%%%%%%%%%%%%%%%%%%%

Let us now show that 
the construction of the previous section gives rise to the 
VEVs for the fields $\phi^a$ of the desired form.
We assume a homogeneous and isotropic Ansatz with spatially flat
metric allowing for a general cosmological evolution,
\be
\label{FRWAns}
N(t)~,~~~~\gamma_{ij}=a^2(t)\delta_{ij}~,
~~~~V^i_a=\frac{M_V}{a(t)}\delta^i_a~,~~~~\phi^a=\Phi(t)x^a\;.
\ee
Substituting this Ansatz into the equations of motion
obtained from putting together the actions ~\eqref{khronoact},
\eqref{phiact}, \eqref{vectact} and \eqref{SVphi}  
we obtain,\footnote{The simplest way to
  derive these equations is to substitute the Ansatz~\eqref{FRWAns} into the
   action and perform variation with respect to the free functions
   $N(t)$ and $a(t)$
   afterwards. 
 Note, however, that it would be incorrect to vary this action 
 with respect to $\Phi$ as the corresponding variation 
$\delta\phi^a=\delta \Phi x^a$ would not be bounded at spatial infinity.}
\bseq
\label{cosmeq}
\begin{align}
\label{cosmeq1}
&3M_c^2H^2-\frac{3\Phi^2}{2a(t)^2}+\frac{3\m^2\Phi}{a(t)}
-{\cal V}_0=\rho_{mat}\;,\\
\label{cosmeq2}
&2 M_c^2\dot H+3M_c^2H^2-\frac{\Phi^2}{2a(t)^2}+\frac{2\m^2\Phi}{a(t)}
-{\cal V}_0=-p_{mat}\;,
\end{align}
\eseq
where 
\be
\label{munew}
\m^2=m_{A}M_V
\ee
and 
\be
\label{Mc}
M_c^2\equiv M_P^2\bigg(1+\frac{\b+3\l}{2}\bigg)-\frac{M_V^2}{2},
\ee
is the ``cosmological Planck mass''.
In Eqs.~(\ref{cosmeq})
we fixed the gauge $N=1$ and assumed that the universe is filled
with matter with energy density  $\rho_{mat}$  and pressure $p_{mat}$.
Taking the derivative of (\ref{cosmeq1}) and using the energy
conservation in the matter sector,
\be
\label{econs}
\dot\rho_{mat}+3H(\rho_{mat}+p_{mat})=0\;,
\ee 
we obtain the following equation for $\Phi$,
\be
\label{Phieq}
\dot\Phi(\Phi-\m^2a(t))=0\;.
\ee
This has two branches of solutions. On the branch $\Phi=const$ 
the VEVs of the St\"uckelberg fields actually disappear with
time. Indeed, the invariant
quantity $\gamma^{ij}\d_i\phi^a\d_j\phi^a=3\Phi^2/a(t)^2$ decreases as
the universe expands. 
Besides, we will see shortly that this branch is unstable at late
times 
whenever 
$\m\neq 0$.
The other branch is 
\be
\label{goodbranch}
\Phi=\m^2a(t)\;.
\ee
It corresponds to constant
strength of the St\"uckelberg fields' gradients and is stable. 
In this latter case the cosmological equations (\ref{cosmeq}) reduce
to the  form, 
\bseq
\label{cosmeqstand}
\begin{align}
\label{cosmeqstand1}
&3M_c^2H^2=\rho_{mat}+{\cal V}_0-\frac{3\mu^4}{2}\;,\\
\label{cosmeqstand2}
&2 M_c^2\dot H+3M_c^2H^2=-p_{mat}+{\cal V}_0-\frac{3\mu^4}{2}\;.
\end{align}
\eseq
We see that in this phase $\mu$ produces a negative shift of the
cosmological constant. 
In what follows we will assume that this contribution is canceled by 
the bare cosmological constant,
\be
\label{eq:Mink}
{\cal V}_0=\frac{3\mu^4}{2}\;,
\ee
so that the Minkowski space-time is a solution in the absence of matter. This
is just the usual fine-tuning of the
cosmological constant.

%%%%%%%%%%%%%%%%%%%%%%%%%%%%%%%%
\section{Hierarchy of EFTs and the graviton mass}
%%%%%%%%%%%%%%%%%%%%%%%%%%%%%%%%
\label{sec:hier}

%%%%%%%%%%%%%%%%%%%%%%%%%%%%%%%%
\subsection{Phases with massive gravitons}
%%%%%%%%%%%%%%%%%%%%%%%%%%%%%%%%

\begin{figure}
\begin{center}
\includegraphics[width=0.8\textwidth]{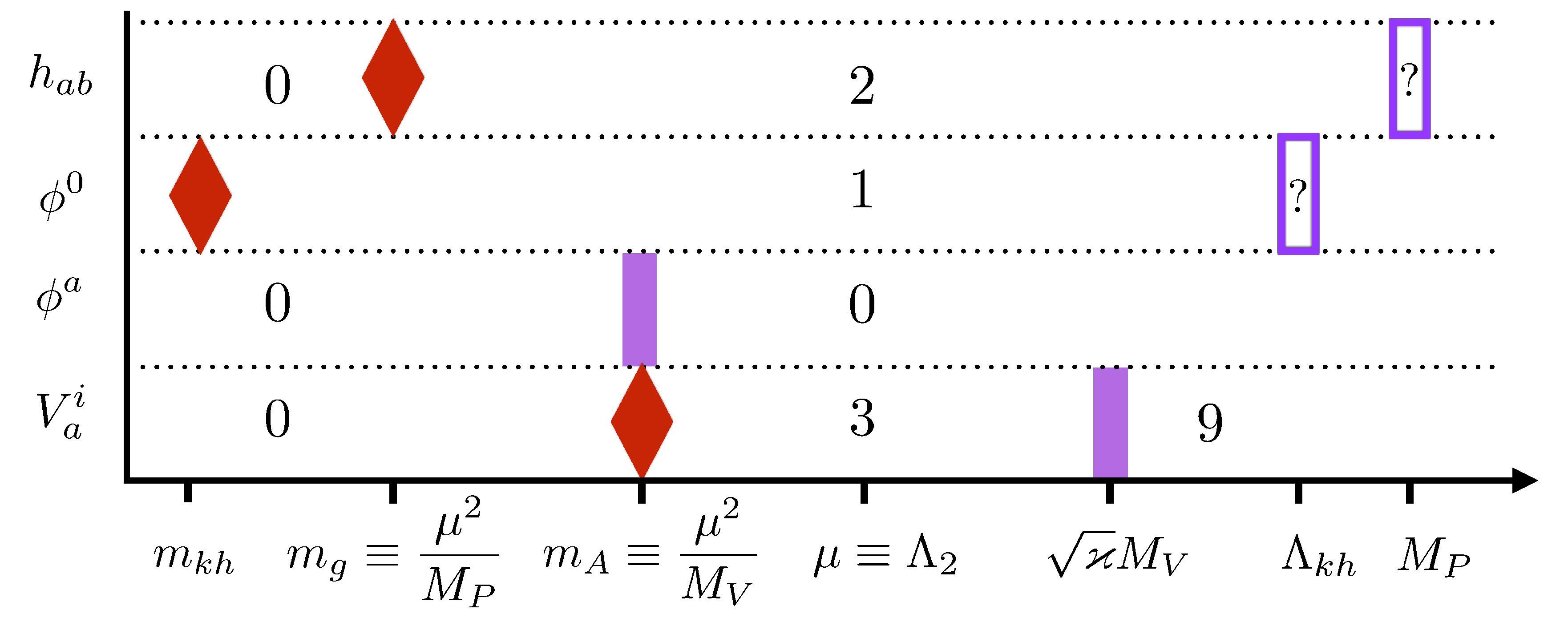}
\caption{\label{fig1}
Relevant energy scales in the theory and the number of propagating
 degrees of freedom in each 
sector at different scales. Rectangles represent the energy scales
at which a sector gets UV completed (we do not make any assumptions 
about the UV completion of the khronon and spin-2 sectors,  
but Ho\v rava gravity \cite{Horava:2009uw} would be a natural
option). 
Rhomboids mark the scales below which a sector loses all its
propagating degrees of freedom. 
The khronon mass $m_{kh}$ will be derived in Sec.~\ref{sec:Newton}
(see Eq.~(\ref{mkh})).
Note that nothing happens at the scale
$\m\equiv \Lambda_2$ which set the cutoff in the original EFT
formulation of massive gravity.} 
\end{center}
\end{figure}

To understand the effect of the mixing term (\ref{SVphi}) on the
spectrum of the theory let us study small perturbations. As before, we
work in the ADM gauge 
and first focus on the phase with a vacuum from the branch of
solutions (\ref{goodbranch}). 
To simplify the analysis, we freeze out the perturbations in the
khronometric sector by sending 
$M_P$ and  $\Lambda_{kh}$ to infinity while keeping $M_V$ and $m_A$ 
finite. For the perturbations of the vectors
and the St\"uckelberg fields we write,
\be
\label{Vphipert}
V^i_a=M_V\delta^i_a+v^i_a~,~~~\phi^a=\m^2 x^a+\psi^a\;.
\ee
If we are interested in energies below $\sqrt{\vk}M_V$ we can 
adopt the $\sigma$-model description. Inserting (\ref{Vphipert})
in 
the constraint equation (\ref{constr}) yields,
\be
\label{vA1}
v^i_a=A^i_a-\frac{A^j_iA^j_a}{2M_V}+O(A^3)\;,
\ee
where $A^i_a$ is an antisymmetric matrix,
$A^i_a+A^a_i=0$. Substituting this into the Lagrangian we obtain, 
\be
\label{massL1}
{\cal L}_\phi+{\cal L}_{V\phi}=-\frac{(\d_i\psi^a)^2}{2}
-\frac{m_A^2}{2}A^j_aA^j_a+m_A A^i_a\d_i\psi^a\;.
\ee
The second term gives mass of
order $m_A$ to the antisymmetric perturbations $A^i_a$.
Below this scale the perturbations of the vectors can be integrated
out completely. From
(\ref{massL1}) we find
\be
\label{Apsi}
A^i_a=\frac{1}{2m_A}(\d_i\psi^a-\d_a\psi^i)\;,
\ee 
which substituted back into (\ref{massL1}) gives (up to a total derivative)
\be
\label{massL2}
{\cal L}_\phi+{\cal L}_{V\phi}=-\frac{\d_i\psi^a\d_i\psi^a }{4}
-\frac{(\d_a\psi^a)^2}{4}\;.
\ee
This coincides with the third and fourth terms (with $\kappa=0$) of 
the quadratic St\"uckelberg Lagrangian (\ref{psi1}) arising in
massive gravity. The ``massless'' fields $\psi^a$ can be interpreted 
as the Goldstone bosons for the broken symmetries 
$\mathit{FDiff}\times SO(3)\to SO(3)_{\mathrm{diag}}$. 
Recall that since we are dealing with LV theories, 
the counting and properties of such fields are different from  
the Lorentz invariant case, see e.g.  
\cite{Brauner:2010wm,Nicolis:2012vf,Watanabe:2012hr}. 
In the same spirit, the vector fields $A^i_a$ that have been
integrated out can be interpreted as the ``Higgs'' fields regularising
the bad behaviour of the Goldstone sector at energies above $m_A$. 
Given the previous results,  we expect that the graviton in this model will
acquire the mass (\ref{gravmass}). For the case $M_P\gg M_V$ the vector
and graviton masses are well separated and at energies $m_A\gg E\gg
m_g$ the dynamics is well described by the EFT for the St\"uckelberg
fields. The hierarchy of various scales in the theory and the
corresponding EFT descriptions are summarized in Fig.~\ref{fig1}.

Alternatively, we can work in the unitary gauge and fix
$\psi^a=0$ at the expense of allowing for the fluctuations of the metric 
\be
\label{metrpert}
N=1+n~,~~~N^i~,~~~\gamma_{ij}=\delta_{ij}+h_{ij}\;.
\ee
The relevant part of the Lagrangian takes the form,
\be
\label{massL3}
{\cal L}_\phi+{\cal L}_{V\phi}=\m^4\bigg[-\frac{\gamma^{aa}}{2}
+\frac{V^a_a}{M_V}-\frac{3}{2}\bigg]\;.
\ee
The solution of the constraint (\ref{constr}) now reads,
\be
\label{vA2}
v^i_a=A^i_a-\frac{M_V}{2}h_{ai}-\frac{A^j_iA^j_a}{2M_V}-
\frac{A^j_i h_{ja}}{4}-\frac{A^j_a h_{ji}}{4}
+\frac{3M_V}{8}h_{ji}h_{ja}+O(A^3,h^3)\;.
\ee
Substituting this formula and the expression
\be
\label{inversegamma}
\gamma^{ij}=\delta_{ij}-h_{ij}+h_{ik}h_{kj}+O(h^3)
\ee
into (\ref{massL3}) we obtain at the quadratic level,
\be
\label{massL4}
{\cal L}_\phi+{\cal L}_{V\phi}=-\frac{m_A^2}{2}A^i_aA^i_a
-\frac{\m^4}{8}h_{ia}h_{ia}\;.
\ee
The first term again gives mass to the antisymmetric perturbations,
while the second explicitly provides the mass term for helicity-2
graviton. As we are going to see in Sec.~\ref{sec:Newton}, it also
gives mass to the khronon (see Eq.~(\ref{mkh})). 
Note that we obtain only one of the two structures for the metric mass term
allowed by the symmetries, cf. (\ref{h11}). This a consequence of
the assumption $M_V\gg m_A$ which implies that the symmetric part of
the vector fluctuations is much heavier (with the mass of order
$\sqrt{\vk}M_V$) than the antisymmetric part. This renders the
parameter $\kappa$ in (\ref{h11}) suppressed by the ratio $m_A^2/\vk
M_V^2$ which we neglected in the above analysis. Were we to make a
different assumption $M_V\sim m_A$, we would obtain both terms of
(\ref{h11}) with comparable coefficients. Finally, if instead of the
khronometric setting one used the ghost condensate for the
$\phi^0$-sector, which in the ADM gauge amounts to promoting all
couplings in the action to functions of the lapse $N$
\cite{Blas:2010hb}, one would be able to reproduce also the other terms
in the general Lagrangian (\ref{h1}) of the massive gravity discussed in
Sec.~\ref{sec:LV}. 

%%%%%%%%%%%%%%%%%%%%%%%%%%%%%%%%
\subsection{Other phases?}\label{sec:otherp}
%%%%%%%%%%%%%%%%%%%%%%%%%%%%%%%%

In the previous section we focused on the branch (\ref{goodbranch})
of the background solutions. 
However, as noticed before, 
the equation (\ref{Phieq}) also admits a second branch
\be
\label{eq:vac2}
\dot \Phi=0\;.
\ee
On this branch the effect of the St\"uckelberg gradients (if non-zero
initially) decays with time in an expanding universe. For completeness
we now analyze the small perturbations around this branch. We write,
\be
\label{phipert1}
\phi^a=\Phi x^a+\psi^a\;,
\ee
with $\Phi=const$ and take the Friedmann--Robertson--Walker (FRW) form for
the metric. We again work in the decoupling limit
$M_P,\Lambda_{kh}\to\infty$, so that the metric fluctuations are
frozen. Below the scale $\sqrt{\vk} M_V$ the fluctuations of the
vectors are restricted to the antisymmetric part $A_a^i$. Expanding
the relevant part of the action to quadratic order we obtain, 
\be
\label{massL5}
\begin{split}
&S_V+S_\phi+S_{V\phi}=\\
&\int dt\, d^3 x
\bigg[\frac{a^3}{2}(\dot A^i_a)^2-\frac{a}{2}\Big(c_1^2(\d_i A^j_a)^2+c_2^2(\d_i
A^i_a)^2+(\d_i\psi^a)^2\Big)
-\frac{a^2 m_A\Phi}{2M_V}(A^i_a)^2+a^2m_A A^i_a\d_i\psi^a\bigg]\;.
\end{split}
\ee
Restricting to the modes with frequencies much higher than the Hubble
rate, we can neglect the terms with derivatives of the scale factor in
the equations of motion. This yields,
\bseq
\label{Aphi}
\begin{align}
\label{Aphi1}
&-{\ddot{A}}^i_a+\frac{c_1^2}{a^2}\d_j^2 A^i_a
+\frac{c_2^2}{a^2}\pd_j \pd_{[i}A^j_{a]}-\frac{m_A\Phi}{M_V a}A^i_a
+\frac{m_A}{a}\d_{[i}\psi^{a]}=0\;,\\
\label{Aphi2}
&\d_i^2\psi^a-am_A\d_iA^i_a=0\;,
\end{align}
\eseq
where the square brackets stand for the antisymmetrization of
indices. Let us perform the 
Fourier transform and concentrate on the transverse modes
\be
\psi^a=e_a^{(\a)}\psi_{(\a)}\;,~~~~
A^i_a=\frac{\bar k_i e_a^{(\a)}-\bar k_ae_i^{(\a)}}{\bar k} A_{(\a)}\;,
\ee
where $e_i^{(\a)}$, $\a=1,2$, are unit polarization vectors orthogonal to the
3-momentum $\bar k_i$. Substituting this into Eqs.~(\ref{Aphi}) and
eliminating $\psi_{(\a)}$ we obtain,
\be
\bigg[\omega^2-\left(c_1^2+\frac{c_2^2}{2}\right)\frac{\bar k^2}{a^2}
-\frac{m_A}{M_V}\bigg(\frac{\Phi}{a}-\frac{\m^2}{2}\bigg)
\bigg]A_{(\a)}=0\;,
\ee
where we used $\m^2$ defined in (\ref{munew}).
We see that whenever $\Phi<\m^2a/2$ the mode is
tachyonic. In particular, the trivial configuration of the
St\"uckelberg fields 
$\phi^a=0$ is unstable. 
Furthermore, in an expanding universe $\m^2a/2$ will exceed any
constant value of $\Phi$ and the instability will set
in at late times. Thus we conclude that in an expanding universe
 this branch is
unstable and we do not consider it any more in this paper.

%%%%%%%%%%%%%%%%%%%%%%%%%%%%%%%
\subsection{Quantum treatment of instantaneous modes}
\label{sec:quantum}   
%%%%%%%%%%%%%%%%%%%%%%%%%%%%%%%

We have argued above that the constructed model is a valid quantum
theory up to the scale (\ref{cutoffkh}). 
We have based this claim on the scaling argument borrowed from
relativistic theories, so it is worth taking a closer look at it to
check if it is not spoiled by Lorentz violation. To get a flavor of
the potential problems consider the instantaneous modes $\phi^a$. 
Let us first switch off their mixing with the vectors by setting $m_A=0$
and perform their perturbative quantization using the path integral
formalism. From 
\eqref{phiact} one reads off the propagator,
\be
\label{propag}
\begin{fmffile}{philineab}
\parbox{70pt}{\begin{fmfgraph*}(50,30)
\fmfpen{thick}
\fmfleft{i} 
\fmfright{o}
\fmflabel{$\phi^a$}{i}
\fmflabel{$\phi^b$}{o}
\fmf{plain,label=$p$,label.side=left}{i,o}
\end{fmfgraph*}}
\end{fmffile}
=-\frac{i}{\bar p^2}\delta^{ab},
\ee
where we denoted by the bar the spatial part of a four-vector 
$p_\m=(p_0,\bar p_i)$.
This propagator 
does not depend on the frequency $p_0$. 
The fields $\phi^a$ couple to the metric and contribute into the
effective action for the perturbations $h_{ij}$. For example,
the one-loop contribution into the quadratic part is 
\be
\label{loop1}
\begin{fmffile}{loop1}
\parbox{110pt}{\begin{fmfgraph*}(80,80)
\fmfpen{thick}
\fmfleft{i} 
\fmfright{o}
\fmfdot{i}
\fmfdot{o}
\fmflabel{$h^{ij}$}{i}
\fmflabel{$h^{kl}$}{o}
\fmf{plain,label=$p+q$,label.side=left,left=0.7,tension=1/3}{i,o}
\fmf{plain,label=$q$,label.side=left,left=0.7,tension=1/3}{o,i}
\end{fmfgraph*}}
\end{fmffile}
=\frac{1}{4}h^{ij}(p)h^{kl}(-p)\int \frac{d q_0}{2\pi}
\int \frac{d^3\bar q}{(2\pi)^3}\;
\frac{\bar q_i\bar q_k(\bar q+\bar p)_j(\bar q+\bar p)_l}
{\bar q^2(\bar q+\bar p)^2}.
\ee
This expression contains two types of divergences. 
The integral over the spatial momentum can be regulated 
by subtracting a finite number of local counterterms. 
However, the whole contribution will still be infinite because of the
overall divergent integral over $q_0$. Note that this divergence has
non-polynomial dependence on the external spatial momentum $\bar p_i$
and therefore is  
{\em non-local in space}.
On the other hand, it does not depend on $p_0$ and hence is 
{\em local in time}.
Thus it can be regulated by introducing a spatially 
non-local counterterm in the bare action. 
Though
unusual, such counterterms do not spoil the validity of the theory. 
In particular, the diagram (\ref{loop1}) does
not contain any imaginary 
part and thus the corresponding counterterm does not violate
unitarity.

One may object that allowing for non-locality, even restricted to only
spatial dimensions, introduces an infinite freedom in the choice of
the bare action. However, we now argue that there is a natural
choice of counterterms for the diagrams where, like in (\ref{loop1}),
a divergent integral over the loop frequency completely factors out of
a frequency-independent part. This consists in canceling these diagrams
altogether. In the present case this would mean that all loop diagrams
containing the instantaneous fields $\phi^a$ must be put to zero. Two
arguments support this prescription. 
First, the integrals over frequency
diverge linearly and thus vanish identically in dimensional
regularization. 
Second, 
we can appeal to the canonical quantization. 
In this formalism, the fields
$\phi^a$ are subject to second-class constraints which force them to
vanish. Indeed, as the action does not depend on the time-derivative
of these fields, the canonical momenta conjugate to them vanish
trivially, while the fields themselves obey the equation,
\be
\label{Laplace}
\nabla_i(N\nabla^i\phi^a)=0\;.
\ee
Supplemented by the vanishing boundary conditions at spatial infinity
it forces\footnote{Multiplying (\ref{Laplace}) by $\phi^a$ and
  integrating over the three-dimensional space we obtain,
\[
0=\int d^3x\, \phi^a\nabla_i(N\nabla^i\phi^a)=
-\int d^3x\, N\nabla_i\phi^a\nabla^i\phi^a\;.
\]
As the lapse function is non-zero everywhere, one concludes that
$\nabla_i\phi^a=0$ and hence $\phi^a$ vanishes due to the boundary conditions. 
}
$\phi^a=0$. In the canonical approach such constrained
degrees of freedom must be eliminated from the start, even prior to
quantization, implying that they completely drop off from the quantum
theory\footnote{There is no modification of the canonical
  structure for the remaining fields as in the case at hand Dirac and
  Poisson brackets are identical.}.

There is a way to implement the above prescription 
within the path integral approach without introducing non-local
counterterms from the start. One notices that the overall result of
integration over $\phi^a$ is a factor
\be
\label{detphi}
\big[\det(i\gamma^{ij}\d_i\d_j)\big]^{-3/2}
\ee
in the partition function. This can be canceled by adding to the
system three real bosonic fields 
$\tilde \phi^a$ and three complex fermionic fields $\eta^a$
with the action,
\be
\label{phitilde}
S_{\tilde \phi\eta}=\int dt\, d^3x\sqrt\gamma N\;\bigg[-\frac{1}{2}
\gamma^{ij}\d_i\tilde\phi^a\d_j\tilde\phi^a
-\gamma^{ij}\d_i\eta^a\d_j\bar\eta^a\bigg]\;.
\ee  
Integrating out these ``remover'' fields multiplies the partition
function by,
\be
\label{detphitilde}
\frac{\big[\det(i\gamma^{ij}\d_i\d_j)\big]^3}
{\big[\det(i\gamma^{ij}\d_i\d_j)\big]^{3/2}}
=\big[\det(i\gamma^{ij}\d_i\d_j)\big]^{3/2}\;,
\ee
which precisely compensates (\ref{detphi}). The expression
(\ref{detphitilde}) corresponds to the 
spatially non-local counterterms discussed above.

Turning on $m_A$ makes the situation less trivial. However, given that
the mixing (\ref{SVphi}) is a 
relevant deformation it clearly cannot spoil
the UV consistency of 
the theory.
A comprehensive analysis of the quantum properties of the theory
introduced in Sec.~\ref{sec:beyond} is beyond the scope of this
paper. Instead, we illustrate the expected behavior in a 
toy model containing a scalar and a vector without any VEVs
in an external non-dynamical metric (we assume $N_i=0$),
\be
\label{toymod}
S=\int dt d^3x\sqrt{\gamma} N\bigg[\frac{\gamma_{ij}\dot V^i\dot V^j}{2N^2}
-\frac{\gamma_{jk}\nabla_iV^j\nabla^i V^k}{2}-\frac{1}{2}\gamma^{ij}\d_i\phi\d_j\phi
+m_A V^i\d_i\phi-\frac{M_V^2}{2}\gamma_{ij}V^iV^j\bigg]\;.
\ee
For simplicity, we have retained only one of the gradient terms for the
vector putting the coefficient in front of it to $c_1^2=1$. 
As before, there are two ways to proceed. In the canonical approach we
have to solve for the field $\phi$ before quantization, 
\be
\label{tpyphi}
\phi=\frac{m_A (\nabla_i V^i+a_i V^i)}{\gamma^{kl}\nabla_k\nabla_l+a^l\nabla_l}\;,
\ee
where 
\be
\label{ai}
a_i\equiv N^{-1}\d_i N\;.
\ee
Substituting this into (\ref{toymod}) we obtain a non-local action which depends
only on $V^i$,
\be
\label{toymod1}
\begin{split}
S=\int dt\, d^3x\sqrt{\gamma} N\bigg[
\frac{\gamma_{ij}\dot V^i\dot V^j}{2N^2}
&-\frac{\gamma_{jk}\nabla_iV^j\nabla^i V^k}{2}
-\frac{M_V^2\gamma_{ij}V^iV^j}{2}\\
&-(\nabla_i V^i+a_i V^i)\frac{m_A^2}{2(\gamma^{kl}\nabla_k\nabla_l+a^k\nabla_k)}
(\nabla_j V^j+a_j V^j)\bigg]\,.
\end{split}
\ee
The Dirac bracket remains  identical to the canonical commutator. One
observes that the limit 
$m_A\to 0$ is smooth and corresponds to restoration of locality.
As non-locality is purely spatial, it does not, in principle, present an
obstruction to canonical quantization. 

However, in practice it is very inconvenient to work 
with the non-local action (\ref{toymod1}). It is more efficient 
to use the path integral approach and retain $\phi$ as a
quantum field.
Assuming that the metric is close to flat, we obtain from
(\ref{toymod}) the propagators
for $\phi$ and $V^i$,
\bseq
\label{toyprops}
\begin{align}
\label{toyprop1}
&\begin{fmffile}{philine}
\parbox{70pt}{\begin{fmfgraph*}(50,30)
\fmfpen{thick}
\fmfleft{i} 
\fmfright{o}
\fmflabel{$\phi$}{i}
\fmflabel{$\phi$}{o}
\fmf{plain,label=$p$,label.side=left}{i,o}
\end{fmfgraph*}}
\end{fmffile}
=  
-\frac{i}{\bar p^2}+\frac{i m_A^2}{\bar p^2(p_0^2-\bar p^2-M_V^2+m_A^2)}\;,\\
\label{toyprop2}
&\begin{fmffile}{phiVline}
\parbox{70pt}{\begin{fmfgraph*}(50,30)
\fmfpen{thick}
\fmfleft{i} 
\fmfright{o}
\fmflabel{$\phi$}{i}
\fmflabel{$V^{i}$}{o}
\fmf{dashes_arrow,label=$p$,label.side=left}{i,o}
\end{fmfgraph*}}
\end{fmffile}
= \frac{-m_A \bar p_i}{\bar p^2(p_0^2-\bar p^2-M_V^2+m_A^2)}\;,\\
\label{toyprop3}
&\begin{fmffile}{Vline}
\parbox{70pt}{\begin{fmfgraph*}(50,30)
\fmfpen{thick}
\fmfleft{i} 
\fmfright{o}
\fmflabel{$V^{i}$}{i}
\fmflabel{$V^{j}$}{o}
\fmf{wiggly,label=$p$,label.side=left}{i,o}
\end{fmfgraph*}}
\end{fmffile}
=\bigg(\delta_{ij}-\frac{\bar p_i \bar p_j}{\bar p^2}\bigg)
\frac{i}{\bar p_0^2-\bar p^2-M_V^2}
+\frac{\bar p_i\bar p_j}{\bar p^2}
\frac{i}{p_0^2-\bar p^2-M_V^2+m_A^2}\;.
\end{align}
\eseq
To avoid cluttered formulas, we will set $M_V=m_A$ in what
follows. This does not affect the UV properties of the theory. 
Consider again the diagram (\ref{loop1}). Now it contains three
contributions. The first one comes from the product of the first
terms in the propagator (\ref{toyprop1}) and, as before, is eliminated
by adding to the path integral 
the fields $\tilde \phi$, $\eta$
with the action \eqref{phitilde}. Besides, there are contributions
coming from the cross-product of the two terms in (\ref{toyprop1}),
\be
\label{toycross1}
-\frac{m_A^2}{2}h^{ij}(p)h^{kl}(-p)\int\frac{dq_0
  d^3\bar q}{(2\pi)^4}\;
\frac{\bar q_i \bar q_k(q+\bar p)_j(\bar q+\bar p)_l}{\bar q^2(q_0^2-\bar q^2)(\bar q+\bar p)^2}\;,
\ee
as well as from the
square of the second term,
\be
\label{toysquare}
\frac{m_A^4}{4}h^{ij}(p)h^{kl}(-p)\int\frac{dq_0
  d^3\bar q}{(2\pi)^4}\;
\frac{\bar q_i \bar q_k(\bar q+\bar p)_j(\bar q+\bar p)_l}
{\bar q^2(q_0^2-\bar q^2)(\bar q+\bar p)^2((q_0+p_0)^2-(\bar q+\bar p)^2)}\;.
\ee
The divergences in these expressions can be removed by genuinely local
counterterms. Consider, for example, Eq.~(\ref{toycross1}). Introducing
Feynman parameters we obtain,  
\be
\label{toycross2}
%\begin{split}
\int\frac{dq_0
  d^3\bar q}{(2\pi)^4}\;
\frac{\bar q_i\bar q_k(\bar q+\bar p)_j(\bar q+\bar p)_l}{\bar q^2(q_0^2-\bar q^2)(q+p)^2}
=2\int_0^1 \frac{dx_1}{\sqrt{x_1}}\int_0^{1-x_1}\!\!\!dx_2
\int\frac{dq_0' d^3\bar q}{(2\pi)^4}
\frac{\bar q_i\bar q_k(\bar q+\bar p)_j(\bar q+\bar p)_l}{({q_0'}^2-\bar q^2-2\bar q\bar px_2-\bar p^2x_2)^3}\;,
%\end{split}
\ee 
where in the last integral we rescaled the loop frequency. The
integral over the four-momentum on the r.h.s. 
has the standard form and its divergent part is a
polynomial in momenta $\bar p$. It is straightforward to check that 
the integration over Feynman
parameters does not contain any additional divergences.
Thus, we
conclude that the overall divergence of (\ref{toycross2}) is
local both in time and space.
Similar reasoning applies to (\ref{toysquare}).

One may worry that a divergence in the Feynman parameters can appear
in the diagrams that contain the loop frequency in the numerator of
the integrand, because then more powers of the Feynman parameters
descend into the denominator. Let us show that this does not
happen. Consider
the diagram arising from the interactions given by the first and
the third
terms in (\ref{toymod}),
\be
\label{loop2}
~~~~~\begin{fmffile}{loop2}
\parbox{110pt}{\begin{fmfgraph*}(80,80)
\fmfpen{thick}
\fmfleft{i} 
\fmfright{o}
\fmfdot{i}
\fmfdot{o}
\fmflabel{$h_{ij}$}{i}
\fmflabel{$h^{kl}$}{o}
\fmf{dashes_arrow,label=$p+q$,label.side=left,left=0.7,tension=1/3}{i,o}
\fmf{dashes_arrow,label=$q$,label.side=left,left=0.7,tension=1/3}{o,i}
\end{fmfgraph*}}
\end{fmffile}
\!\!\!\!\!=\frac{m_A^2}{2}h_{ij}(p)h^{kl}(\!-p)\!\!
\int \!\!\frac{dq_0d^3\bar q}{(2\pi)^4}\;
\frac{q_0(q_0+p_0)\bar q_i\bar q_k(\bar q+\bar p)_j(\bar q+\bar p)_l}
{\bar q^2(q_0^2\!-\!\bar q^2)(\bar q\!+\!\bar p)^2
((q_0\!+\! p_0)^2\!-\!(\bar q\!+\!\bar p)^2)}\,.
\ee
Passing to the Feynman parameterization we obtain,
\be
\begin{split}
&\int \frac{dq_0d^3\bar q}{(2\pi)^4}\;
\frac{q_0(q_0+p_0)\bar q_i\bar q_k(\bar q+\bar p)_j(\bar q+\bar p)_l}{\bar q^2(q_0^2-\bar q^2)(\bar q+\bar p)^2
((q_0+p_0)^2-(\bar q+\bar p)^2)}\\
&\quad=6\int_0^1 dx_1\int_0^{1-x_1}dx_2\int_0^{1-x_1-x_2}dx_3\\
&\quad\quad\times\int \frac{dq_0 d^3\bar q}{(2\pi)^4}
\frac{q_0(q_0+p_0)\bar q_i\bar q_k(\bar q+\bar p)_j(\bar q+\bar p)_l}{
\big[(x_1+x_2)q_0^2+2q_0p_0 x_2+p_0^2 x_2
-\bar q^2-2\bar q\bar p (x_2+x_3)-\bar p^2(x_2+x_3) \big]^4}\;.
\end{split}
\ee
Upon rescaling of the loop frequency, $q_0\mapsto
q_0'=q_0/\sqrt{x_1+x_2}$, the most singular contribution in
the integral over Feynman parameters is
proportional to $(x_1+x_2)^{-3/2}$ which is again integrable.
At the heuristic level this can be understood as follows. The
  divergences in the integrals over Feynman parameters are usually
  associated to the infrared (or collinear) divergences, which are
  absent in our case because the original expressions
  (\ref{toycross1}) and (\ref{loop2})
are IR safe. 

\begin{figure}
\begin{center}
\begin{fmffile}{loop3}
\parbox{130pt}{\begin{fmfgraph*}(120,180)
%\fmfpen{thick}
\fmfsurroundn{v}{12}
\fmfdot{v2}
\fmfdot{v6}
%\fmfdot{v10}

%\fmf{curly}{v6,v2}
\fmf{plain,left=0.18,tension=1}{v6,v5,v4,v3,v2}
\fmf{plain,right=0.05,tension=1}{v6,v13}
\fmf{plain,right=0.05,tension=1/8}{v13,v15}
\fmf{phantom,right=0.05,tension=1/4}{v15,v9}
\fmf{phantom,tension=1}{v7,v13}
\fmf{phantom,tension=1/2}{v8,v15}
\fmf{plain,left=0.1,tension=1}{v2,v14}
\fmf{plain,left=0.05,tension=1/8}{v14,v16}
\fmf{phantom,left=0.1,tension=1/4}{v16,v11}
\fmf{phantom,tension=1}{v1,v14}
\fmf{phantom,tension=1/2}{v12,v16}
\fmf{plain,right=0.35,tension=0}{v15,v10,v16}
\fmffreeze

\fmf{curly}{v13,v17}
\fmf{plain,left=0.5,tension=1/3}{v17,v18}
\fmf{plain,right=0.7,tension=1/3}{v17,v18}
\fmf{curly,tension=1.5}{v18,v23}
\fmf{curly,tension=1.5}{v23,v19}
\fmf{plain,left=0.5,tension=1/3}{v19,v20}
\fmf{plain,right=0.7,tension=1/3}{v19,v20}
\fmf{curly}{v20,v14}

\fmf{curly,tension=1.2}{v15,v21}
\fmf{plain,left=0.1,tension=1/4}{v21,v22}
\fmf{plain,right=0.4,tension=1/4}{v21,v22}
\fmf{curly,tension=1.2}{v22,v16}

\fmffreeze

\fmf{plain,tension=1}{v23,v24}
\fmf{curly,tension=3}{v24,v21}
%\fmf{phantom}{v7,v24}
\fmf{plain,tension=1}{v23,v25}
\fmf{curly,tension=3}{v25,v22}
%\fmf{phantom}{v1,v25}
\fmf{plain,right=0.2}{v24,v25}
\end{fmfgraph*}}
\end{fmffile}
\end{center}
\caption{\label{fig:VR} Generic diagram with instantaneous modes and
  gravitons propagating in the loops. Summing it with the diagrams of
  the same topology where the different subsets of the $\phi$-loops are replaced
by those of $\tilde\phi$ and $\eta$ will remove all non-local 
divergences.}
\end{figure}
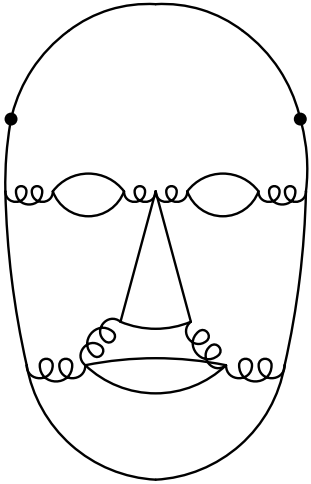

By extending the above reasoning to other diagrams in the model 
 (\ref{toymod}) the reader will easily convince herself that the only
 class of divergences that require (spatially) non-local counterterms
 are those where {\em all} propagators in a given loop are equated to
 the first term in (\ref{toyprop1}). 
These divergences are independent of $m_A$ and are exactly canceled
by the remover fields $\tilde\phi$, $\eta$ with the action
(\ref{phitilde}). Furthermore, 
this cancellation persists upon making the metric 
 $h_{ij}$ dynamical and allowing it to propagate in the loops. Thus, it is
 natural to conjecture that no matter how complicated a diagram is (see an example
 in Fig.~\ref{fig:VR}), it will require only local counterterms after
addition of similar diagrams with the fields $\tilde\phi$ and $\eta$.

So far we have discussed only the instantaneous modes associated with
the St\"uckelberg fields $\phi^a$ of massive gravity. The dynamics of
these fields is relatively simple: they enter in the UV action 
 \eqref{phiact}, \eqref{SVphi} quadratically and do not contain
 any propagating degrees of freedom. This allowed us to eliminate all
 unusual non-local divergences appearing due to these fields by adding
 the ``remover'' sector $\tilde\phi^a$, $\eta^a$ 
with the simple action (\ref{phitilde}).
However, as pointed out in \cite{Blas:2010hb}, another source of
instantaneous interactions is the khronon field $\phi^0$. Here the
situation appears to be more complicated: the khronon describes, besides the
instantaneous mode, a genuine propagating degree of freedom and,
furthermore, enters into the action non-linearly. 
This produces  difficulties with the quantization 
which are intrinsic of Ho\v rava (or khronometric) proposal. 
We plan to address them elsewhere. For now, we just point out that the 
 discussion of this section suggests that 
a consistent quantization of the theory exists.
Indeed, in the decoupling
limit the propagator of the khronon has the form similar to the second
term in (\ref{toyprop1}) \cite{Blas:2010hb}. 
We have seen that the divergences associated
with such propagators can be removed by local counterterms.

%%%%%%%%%%%%%%%%%%%%%%%%%%%%%%%%
\section{Modification of the Newton's law}
\label{sec:Newton}
%%%%%%%%%%%%%%%%%%%%%%%%%%%%%%%%

Having addressed the theoretical consistency of the model, we now
study its immediate
phenomenological consequences. Let us consider 
the gravitational field of a point mass
$M_\odot$ at a fixed position $x^i=0$. 
We will focus on the weak field (linear) regime and assume the minimal
coupling of the metric to the matter sector; the latter is justified
by the phenomenological constraints on deviations from the Lorentz
invariance \cite{Liberati:2013xla}. 
There are two important changes with respect
to the massive gravity phase  \eqref{resdiffeos}  
described in \cite{Dubovsky:2004sg,Dubovsky:2004ud}. 
First, at any energy the role of the St\"uckelberg field $\phi^0$ is
played by the khronon.
Second, the theory is defined also above the energy $\Lambda_2$, 
which can have experimental consequences at short distances relevant 
e.g. in the early universe or in very dense stars.
We will only consider the large distance modification in this section.

We
work in the unitary gauge\footnote{Recall that it is consistent to 
first fix the unitary gauge and take the variation of 
the action afterwards~\cite{Blas:2009yd}.},
 and restrict the vector fields to the coset space \eqref{constr}. 
We consider the scalar part of the
perturbations. The expansion around the Minkowski background to linear order reads,
\bseq
\label{linscal}
\begin{align}
\label{linscla1}
&N=1+\varphi\;,\\
\label{linscla2}
&N_i=\d_iB\;,\\
\label{linscla3}
&\gamma_{ij}=\delta_{ij}-2\bigg(\delta_{ij}-\frac{\d_i\d_j}{\D}\bigg)\Psi
-2\frac{\d_i\d_j}{\D}E\;,\\
\label{linscla4}
&V^i_a=M_V\delta^i_a+\epsilon_{iaj}\d_j C 
+M_V\bigg(\delta_{ia}-\frac{\d_i\d_a}{\D}\bigg)\Psi
+M_V\frac{\d_i\d_a}{\D}E\;,
\end{align}
\eseq
where we have used the linear part of Eq.~(\ref{vA2}). 
Using the expression (\ref{massL4}) and expanding the khronometric and
vector Lagrangians (\ref{khronoact}) and (\ref{vectact}) to quadratic
order one obtains\footnote{We remind that $\m$ is defined in (\ref{munew}).},
\be
\label{L2scal}
\begin{split}
{\cal
  L}^{(2)}_{scal}=&\frac{M_P^2}{2}\bigg[(1-\b)(-2\dot\Psi^2+4\Psi\ddot
E+4\Psi\D\dot B)
-(\l+\b)(2\dot\Psi+\dot E+\D B)^2\\
&-2\Psi\D\Psi+4\varphi\D\Psi+\a(\d_i\varphi)^2\bigg]
+\frac{M_V^2}{2}\Big(2\dot\Psi^2+(\dot E+\D B)^2-4(c_1^2+c_2^2)(\d_i\Psi)^2
\Big)\\
&+(\d_i\dot C)^2-c_1^2(\d_i\d_k C)^2
-\m^4\Psi^2-\frac{\m^4}{2}E^2-m_A^2(\d_i C)^2
-\varphi M_\odot\delta({\bf x})\;.
\end{split}
\ee
We see that the pseudoscalar mode $C$ completely decouples and has the
dispersion relation 
\be
\label{dispa}
\omega^2=c_1^2\bar k^2+m_A^2\;.
\ee
For the other components we obtain the set of equations,
\bseq
\label{scaleqs}
\begin{align}
\label{scaleq1}
&2M_P^2\D\Psi-\a M_P^2\D\varphi-M_\odot\delta({\bf x})=0\;,\\
\label{scaleq2}
&2M_P^2(1+\l)\dot\Psi+\big[M_P^2(\l+\b)-M_V^2\big](\dot E+\D B)=0\;,\\
&\big[M_P^2(1+\b+2\l)-M_V^2\big]\ddot\Psi+M_P^2(1+\l)(\ddot E+\D\dot B)\notag\\
&\qquad\quad~\;-\big[M_P^2-2M_V^2(c_1^2+c_2^2)\big]\Delta \Psi-\m^4\Psi+M_P^2\D\varphi=0\;,
\label{scaleq3}\\
\label{scaleq4}
&2M_P^2(1+\l)\ddot\Psi+\big[M_P^2(\l+\b)-M_V^2\big](\ddot E+\D \dot B)-\m^4 E=0\;.
\end{align}
\eseq
Combining the second and fourth equations we find 
\be
\label{EB}
E=0~,~~~~\D B=-\frac{2M_P^2(1+\l)}{M_P^2(\l+\b)-M_V^2}\dot\Psi\;.
\ee
Substituting this into Eq.~(\ref{scaleq3}) and using (\ref{scaleq1})
to express $\varphi$ we
find the equation for the single variable $\Psi$,
\be
\label{Psieq}
-\a M_P^2\frac{M_P^2(2+3\l-\b)-M_V^2}{M_P^2(\l+\b)-M_V^2}\ddot\Psi
+2M_P^2\bigg(1-\frac{\a}{2}\bigg)\D\Psi-\m^4\a\Psi
=M_\odot\delta^{(3)}({\bf x})\;,
\ee
where we have assumed $\a,\b,\l, M_V/M_P\ll 1$ and kept only up 
to the first subleading order in these parameters.

Let us momentarily put the source to zero, $M_\odot=0$. Then
(\ref{Psieq}) reduces 
to the wave equation for the helicity-0 graviton mode --- the
khronon. To the leading order, its dispersion relation reads,
\be
\label{khronodisp}
\omega^2=\bar k^2\bigg(\frac{\l+\b}{\a}-\frac{M_V^2}{\a M_P^2}\bigg)
+\frac{\m^4}{2M_P^2}\bigg(\l+\b-\frac{M_V^2}{M_P^2}\bigg)\;.
\ee 
One makes two observations. First, the velocity of the khronon,
\be
\label{ckh}
c_{kh}= \sqrt{\frac{\l+\b}{\a}-\frac{M_V^2}{\a M_P^2}},
\ee
gets
renormalized compared to the pure khronometric theory (see
Eq.~(\ref{khronodisp1})) due to the VEVs of the vector fields. The requirement
that the velocity square remains positive puts an upper bound,
\be
\label{vectorbound}
M_V<M_P\sqrt{\l+\b}\;.
\ee
This condition is automatically satisfied within our assumptions
(\ref{hierarchy}). 
Second, the khronon acquires a mass gap
\be
\label{mkh}
m_{kh}=\frac{\m^2}{M_P}\sqrt{\frac{\l+\b}{2}-\frac{M_V^2}{2M_P^2}},
\ee
which is parametrically smaller than the mass of the graviton
(\ref{gravmass}). 
This is
in striking contrast to the case of \cite{Dubovsky:2004sg} where the
St\"uckelberg field $\phi^0$ 
remains massless.
It is worth stressing that the appearance of the gap (\ref{mkh}) is an
IR phenomenon and depends only on the properties of the St\"uckelberg
sector $\phi^0$, $\phi^a$ at energies below $m_A$. Thus one expects it
to be a universal property of massive gravities where this sector
obeys the symmetries (\ref{phiashifts}), (\ref{repar}).

\begin{figure}[tb]
\begin{center}
\includegraphics[width=0.4\textwidth]{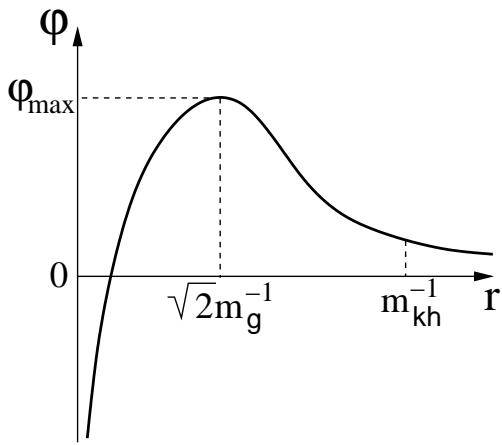}
\caption{\label{fig3}The shape of the Newton potential in the massive
  gravity model of this paper. The gravitational force becomes
  repulsive at distances larger than the inverse graviton mass.} 
\end{center}
\end{figure}

Next, we restore the source in (\ref{Psieq}) and focus on static
configurations. We find,
\be
\label{PsiNewt}
\Psi=-\frac{G_N M_\odot}{r}\e^{-m_{kh}r/c_{kh}}\;,
\ee
where we have introduced the Newton's constant,
\be
\label{GNewt}
G_N\equiv \frac{1}{8\pi M_P^2(1-\a/2)}\;.
\ee
Clearly, the gravitational field has a Yukawa-type
behavior. Finally, from (\ref{scaleq1}) we obtain the Newton's
potential,
\be
\label{phiNewt}
\varphi=-\frac{G_N M_\odot}{r}\bigg[1-\frac{2}{\a}
\Big(1-\e^{-m_{kh}r/c_{kh}}\Big)\bigg]\;.
\ee
This potential is plotted in Fig.~\ref{fig3}.
One observes that it markedly deviates from the
Newtonian potential of general relativity. The most striking feature
is that the gravitational force becomes 
{\em repulsive} at distances
$r>1/m_g$. At large distances the potential goes to zero. This is
different from the case of massive gravities with gapless field
$\phi^0$ \cite{Dubovsky:2004sg}, where the gravitational potential
generically presents linear growth with 
distance\footnote{This growth may be cut by non-linearities of the
  model \cite{Dubovsky:2004sg,Bebronne:2009mz,Comelli:2010bj} 
or by non-stationary evolution of the background
\cite{Blas:2009my}. Also, it is absent if the coefficients in the mass
term (\ref{h1}) satisfy certain relations
\cite{Dubovsky:2004sg,Dubovsky:2004ud}.}
\cite{Dubovsky:2004ud,Blas:2009my}.
Note also that there is no van Dam--Veltman--Zakharov (vDVZ)
discontinuity \cite{vanDam:1970vg,Zakharov:1970cc}: in the limit
$\m\to 0$ the potentials $\varphi$, $\Psi$ reduce to their GR
expressions. 

To understand the behavior of the Newton potential in more
detail, we expand the exponent at $r\ll c_{kh}m_{kh}^{-1}
=(\sqrt{\alpha} m_g)^{-1}$. At these distances the khronon mass is
irrelevant and one expects the potential to coincide with the results
existing in the literature. We obtain,
\be
\label{phiNewt1}
\varphi=G_N M_\odot\bigg[-\frac{1}{r}+\sqrt{\frac{2}{\a}}
m_g-\frac{m_g^2r}{2}+\ldots\bigg]\;,
\ee
where dots
stand for the terms that are suppressed by the powers of the
combination $\sqrt\a m_g r$. The second term in brackets gives a
constant shift of the Newton potential which drops off from the
observables involving only distances $r\lesssim 1/m_g$. The third
term gives precisely the linear contribution discussed in
\cite{Dubovsky:2004ud,Blas:2009my}. 
Note that for our model this contribution is
repulsive. 
The potential reaches a maximum at $r=\sqrt{2}/m_g$
where
\[
\varphi_{max}= \sqrt{2/\a}\,G_N M_\odot m_g\;.
\]  
For the validity
of the linearized approximation $\varphi_{max}$
must be much smaller than one. This translates into the condition
that the graviton mass must be smaller than the inverse Schwarzschild
radius of the source multiplied by $\sqrt\a$.
Unless $\a$ is extremely
small, this condition is not very restrictive.

Stronger 
phenomenological constraints come from the requirement that the 
gravitational field of localized sources 
should not significantly deviate from the standard form at 
astrophysical scales. The Solar System tests put a limit on the
difference between the two gravitational potentials $\varphi$ and
$\Psi$. In the post-Newtonian 
framework this is traditionally parameterized by the ratio
$\gamma\equiv\Psi/\varphi$ and the current constraint 
(measured at the orbit of Saturn by the Cassini satellite) reads
\cite{Will:2014kxa},  
\be
\label{consgam}
\gamma-1=(2.1\pm 2.3)\times 10^{-5}\;.
\ee
From the expressions\footnote{We subtract the constant piece from
  (\ref{phiNewt1}).} (\ref{PsiNewt}), (\ref{phiNewt1}) we obtain the
formula for
$\gamma$ in our model 
at distances shorter than inverse khronon mass,
\be
\label{ourgamma}
\gamma=1-\frac{(m_g r)^2}{2}\;.
\ee
This gives an upper bound 
$m_g<4\times 10^{-17} {\rm cm}^{-1}\sim 120\; \text{pc}^{-1}$.
A tighter limit comes from the gravitational field of
galaxies. The requirement that it matches the standard expression implies
\be
\label{mgconstr}
m_g\lesssim (1~ \text{Mpc})^{-1}\;.
\ee
It is likely that yet stronger bounds can be obtained from the large scale
structure and the cosmic microwave background (CMB). We leave this
analysis for future. 
  
It would be also interesting to explore if the gravitational repulsion
found above can be active at the cosmological scales and lead to
accelerated expansion of the universe. Note that this mechanism of
acceleration would rely crucially on the presence of inhomogeneities,
as the homogeneous FRW Ansatz does not exhibit any self-accelerated
behavior (see Sec.~\ref{sec:cosmology}).  

Before closing this section, let us mention that a complementary way to
constrain the graviton mass is by looking directly at the
modifications in the helicity-2 sector. These have consequences for
radiation and propagation of gravity 
waves \cite{Dubovsky:2004ud,Dubovsky:2009xk,Mirshekari:2011yq}.  
Having a more
complete theory allows to put these studies on the firm ground in the
situations with characteristic scales smaller 
than $\Lambda_2^{-1}$, such as inflation and reheating.

%%%%%%%%%%%%%%%%%%%%%%%%%%%%%%%%
\section{Summary and discussion}
\label{sec:discussion}
%%%%%%%%%%%%%%%%%%%%%%%%%%%%%%%%

In this paper we have proposed an embedding of Lorentz violating
massive gravity above the scale $\Lambda_2\equiv\sqrt{m_gM_P}$. 
The proposed theory has a
high cutoff scale only a few orders of magnitude below the Planck
mass and independent of the mass of the graviton. At high
energies the theory possesses a large symmetry $\mathit{FDiff}\times
SO(3)$ which is {\em spontaneously} broken at lower energy to a
diagonal global $SO(3)$ subgroup\footnote{Notice that Lorentz
  invariance is broken explicitly all the way up to the cutoff.}. 
This pattern of symmetry breaking is
realized by a triplet of space-like vector fields which develop
non-zero VEVs and play the role of the ``Higgs'' fields. 
A crucial technical role is played by a quadratic mixing between the
vectors and the St\"uckelberg fields $\phi^a$ of massive gravity. Once
the vectors acquire VEVs, this mixing forces the St\"uckelbergs to
develop coordinate-dependent profiles, which eventually translates
into the graviton mass. This means that no non-linear interactions in
the St\" uckelberg sector are required to do this job and one can
restrict to purely quadratic action for the fields $\phi^a$, thus
eliminating any strong coupling from this sector. This mechanism is
reminiscent of the proposal for the (partial) UV completion of the
ghost condensate model \cite{Blas:2011en,Ivanov:2014yla} 
where a mixing between a time-like
vector acquiring a VEV and a massless scalar forces the latter to
evolve in time.   

The graviton mass in the model is proportional to the product of the
vector VEVs and the coefficient in front of the vector-St\"uckelberg
mixing. Thus, it vanishes both if the vector VEVs disappear (in the
unbroken phase) or if the mixing is switched off. The action stays
regular in the limit of vanishing mass and therefore one expects all
observable quantities, with the quantum corrections included,
to behave smoothly in this limit. In this sense our mechanism is
analogous to the Higgs mechanism of gauge theories. It is worth
stressing that in our model the mixing between the vector and
St\"uckelberg fields is protected by a discrete symmetry
$\phi^a\mapsto -\phi^a$ and thus a small coefficient in front of it is
technically natural. This implies that the graviton mass is stable
under quantum corrections.   

We analyzed the structure of the theory at different energies and
explicitly verified the expectation that new degrees of freedom,
besides those of pure massive gravity, must exist below the scale
$\Lambda_2$. Indeed, we found that certain components of the vector
fields propagate at these energies. These degrees of freedom have a
mass gap which is parametrically smaller than $\Lambda_2$, but
still bigger than $m_g$. It would be interesting to work out the
consequences of these new light degrees of freedom for phenomenology.

We also found that the helicity-0 component of the graviton, which in our
model is identified with the khronon of the khronometric model,
acquires a mass parametrically lower than $m_g$. This has important
implications for the gravitational potentials of localized sources:
unlike previous models of LV massive gravity, in our case the potentials fall of
exponentially at large distances. Remarkably, the shape of the Newton
potential is not monotonic. It grows from negative values at short
distances, changes sign, reaches a positive maximum at $r=\sqrt{2}m_g^{-1}$
and then decreases towards $r\to\infty$. This implies that the
gravitational force becomes {\em repulsive} at $r>\sqrt{2}m_g^{-1}$. This
property may lead to a rich phenomenology which we leave for
future studies. An interesting question is whether the gravitational
repulsion between the inhomogeneities present in the universe can
provide the accelerated expansion at recent epoch, despite the fact
that for the strictly homogeneous Ansatz our model does not exhibit
any self-acceleration.

A subtle theoretical aspect of our model, inherited from the effective
theory of LV massive gravity, is the presence of instantaneous
interactions. We have addressed the issue of quantization of the
instantaneous modes and argued that it can be performed
consistently. We also pointed out that in the canonical formalism the
instantaneous modes must be interpreted as a certain type of
non-locality along the spatial dimensions. To make the discussion
concise, we focused on simplified toy models. A more comprehensive
study of this topic is definitely required owing to its importance for
LV proposals for quantum gravity \cite{Horava:2009uw,Blas:2010hb}. 

Another open question left for future research is to understand how
the strong coupling of LV massive gravity manifests itself at the
level of Feynman diagrams and how it is canceled by the new degrees of
freedom appearing in our model 
(see
\cite{Schwartz:2003vj,Aubert:2003je} for related works in the Lorentz
invariant context). This may shed light on possible generalizations of
the mechanism proposed in this paper to other IR modifications of
gravity, such as multi-metric theories and the Lorentz invariant setup
of \cite{deRham:2010kj}. In particular, it would be interesting to
prove at the diagrammatic level the 
(im)possibility of a Lorentz invariant Wilsonian UV completion of the
latter setup.

\section*{Acknowledgments}
We are grateful to Denis Comelli, Sergei Dubovsky, Maxim Pospelov and Mikhail
Ivanov for useful discussions. We also thank Claudia de Rham and
Gregory Gabadadze for useful comments on the draft.
S.S. is grateful to the Perimeter Institute for hospitality during this
work.
Research at Perimeter Institute is supported by the Government of
Canada through Industry Canada and by the Province of Ontario through
the Ministry of Economic Development \& Innovation.

\end{document}